\documentclass[twocolumn,appendixfloats,twocolappendix]{aastex631}

\usepackage{natbib}
\usepackage{amsmath,amssymb}
\usepackage{xspace}
\usepackage{rotating}
\usepackage{enumitem}
\usepackage{booktabs}
\usepackage{pifont}
\usepackage{bm} 

\usepackage{macros}
\usepackage{tikz,xcolor,hyperref}
\usepackage{fdsymbol}
\usepackage{graphicx} 
\usepackage{tikz}
\usetikzlibrary{arrows,positioning}
\usetikzlibrary{external}
\definecolor{mpl_blue}{HTML}{1F77B4}
\definecolor{mpl_orange}{HTML}{FF7F0E}
\definecolor{mpl_green}{HTML}{2CA02C}
\definecolor{mpl_red}{HTML}{D62728}

\usepackage[all]{hypcap}
\usepackage[caption=false]{subfig}
\newcommand{\m}[1]{\mathbf{#1}}

\def\be{\begin{equation}}
\def\ee{\end{equation}}
\newcommand{\bb}{\begin{bmatrix}}
\newcommand{\eb}{\end{bmatrix}}
\def\bea{\begin{eqnarray}}
\def\eea{\end{eqnarray}}

\def\R{\color{red}}

\defcitealias{15yrGWB}{NG15}
\defcitealias{12.5yrGWB}{NG12.5}
\defcitealias{11yrGWB}{NG11}
\defcitealias{9yrGWB}{NG9}

\begin{document}
\title{The NANOGrav 15 yr Data Set: Harmonic Analysis of the Pulsar Angular Correlations}

\shorttitle{NANOGrav 15 yr Harmonic Analysis}
\shortauthors{The NANOGrav Collaboration}

\author[0000-0001-5134-3925]{Gabriella Agazie}
\affiliation{Center for Gravitation, Cosmology and Astrophysics, Department of Physics, University of Wisconsin-Milwaukee,\\ P.O. Box 413, Milwaukee, WI 53201, USA}

\author[0000-0002-4972-1525]{Jeremy G. Baier}
\affiliation{Department of Physics, Oregon State University, Corvallis, OR 97331, USA}

\author[0000-0003-2745-753X]{Paul T. Baker}
\affiliation{Department of Physics and Astronomy, Widener University, One University Place, Chester, PA 19013, USA}

\author[0000-0003-0909-5563]{Bence B\'{e}csy}
\affiliation{Department of Physics, Oregon State University, Corvallis, OR 97331, USA}

\author[0000-0002-2183-1087]{Laura Blecha}
\affiliation{Physics Department, University of Florida, Gainesville, FL 32611, USA}

\author[0000-0003-1928-4667]{Kimberly K. Boddy}
\affiliation{Department of Physics, The University of Texas at Austin, Austin, TX 78712, USA}

\author[0000-0001-6341-7178]{Adam Brazier}
\affiliation{Cornell Center for Astrophysics and Planetary Science and Department of Astronomy, Cornell University, Ithaca, NY 14853, USA}
\affiliation{Cornell Center for Advanced Computing, Cornell University, Ithaca, NY 14853, USA}

\author[0000-0003-3053-6538]{Paul R. Brook}
\affiliation{Institute for Gravitational Wave Astronomy and School of Physics and Astronomy, University of Birmingham, Edgbaston, Birmingham B15 2TT, UK}

\author[0000-0003-4052-7838]{Sarah Burke-Spolaor}
\altaffiliation{Sloan Fellow}
\affiliation{Department of Physics and Astronomy, West Virginia University, P.O. Box 6315, Morgantown, WV 26506, USA}
\affiliation{Center for Gravitational Waves and Cosmology, West Virginia University, Chestnut Ridge Research Building, Morgantown, WV 26505, USA}

\author{Rand Burnette}
\affiliation{Department of Physics, Oregon State University, Corvallis, OR 97331, USA}

\author[0000-0002-5557-4007]{J. Andrew Casey-Clyde}
\affiliation{Department of Physics, University of Connecticut, 196 Auditorium Road, U-3046, Storrs, CT 06269-3046, USA}

\author[0000-0003-3579-2522]{Maria Charisi}
\affiliation{Department of Physics and Astronomy, Vanderbilt University, 2301 Vanderbilt Place, Nashville, TN 37235, USA}

\author[0000-0002-2878-1502]{Shami Chatterjee}
\affiliation{Cornell Center for Astrophysics and Planetary Science and Department of Astronomy, Cornell University, Ithaca, NY 14853, USA}

\author[0000-0001-7587-5483]{Tyler Cohen}
\affiliation{Department of Physics, New Mexico Institute of Mining and Technology, 801 Leroy Place, Socorro, NM 87801, USA}

\author[0000-0002-4049-1882]{James M. Cordes}
\affiliation{Cornell Center for Astrophysics and Planetary Science and Department of Astronomy, Cornell University, Ithaca, NY 14853, USA}

\author[0000-0002-7435-0869]{Neil J. Cornish}
\affiliation{Department of Physics, Montana State University, Bozeman, MT 59717, USA}

\author[0000-0002-2578-0360]{Fronefield Crawford}
\affiliation{Department of Physics and Astronomy, Franklin \& Marshall College, P.O. Box 3003, Lancaster, PA 17604, USA}

\author[0000-0002-6039-692X]{H. Thankful Cromartie}
\affiliation{National Research Council Research Associate, National Academy of Sciences, Washington, DC 20001, USA resident at Naval Research Laboratory, Washington, DC 20375, USA}

\author[0000-0002-2185-1790]{Megan E. DeCesar}
\affiliation{George Mason University, Fairfax, VA 22030, resident at the U.S. Naval Research Laboratory, Washington, DC 20375, USA}

\author[0000-0002-6664-965X]{Paul B. Demorest}
\affiliation{National Radio Astronomy Observatory, 1003 Lopezville Rd., Socorro, NM 87801, USA}

\author{Heling Deng}
\affiliation{Department of Physics, Oregon State University, Corvallis, OR 97331, USA}

\author[0000-0002-2554-0674]{Lankeswar Dey}
\affiliation{Department of Physics and Astronomy, West Virginia University, P.O. Box 6315, Morgantown, WV 26506, USA}
\affiliation{Center for Gravitational Waves and Cosmology, West Virginia University, Chestnut Ridge Research Building, Morgantown, WV 26505, USA}

\author[0000-0001-8885-6388]{Timothy Dolch}
\affiliation{Department of Physics, Hillsdale College, 33 E. College Street, Hillsdale, MI 49242, USA}
\affiliation{Eureka Scientific, 2452 Delmer Street, Suite 100, Oakland, CA 94602-3017, USA}

\author[0000-0001-7828-7708]{Elizabeth C. Ferrara}
\affiliation{Department of Astronomy, University of Maryland, College Park, MD 20742, USA}
\affiliation{Center for Research and Exploration in Space Science and Technology, NASA/GSFC, Greenbelt, MD 20771}
\affiliation{NASA Goddard Space Flight Center, Greenbelt, MD 20771, USA}

\author[0000-0001-5645-5336]{William Fiore}
\affiliation{Department of Physics and Astronomy, West Virginia University, P.O. Box 6315, Morgantown, WV 26506, USA}
\affiliation{Center for Gravitational Waves and Cosmology, West Virginia University, Chestnut Ridge Research Building, Morgantown, WV 26505, USA}

\author[0000-0001-8384-5049]{Emmanuel Fonseca}
\affiliation{Department of Physics and Astronomy, West Virginia University, P.O. Box 6315, Morgantown, WV 26506, USA}
\affiliation{Center for Gravitational Waves and Cosmology, West Virginia University, Chestnut Ridge Research Building, Morgantown, WV 26505, USA}

\author[0000-0001-7624-4616]{Gabriel E. Freedman}
\affiliation{Center for Gravitation, Cosmology and Astrophysics, Department of Physics, University of Wisconsin-Milwaukee,\\ P.O. Box 413, Milwaukee, WI 53201, USA}

\author[0000-0002-8857-613X]{Emiko C. Gardiner}
\affiliation{Department of Astronomy, University of California, Berkeley, 501 Campbell Hall \#3411, Berkeley, CA 94720, USA}

\author{Kyle A. Gersbach}
\affiliation{Department of Physics and Astronomy, Vanderbilt University, 2301 Vanderbilt Place, Nashville, TN 37235, USA}

\author[0000-0003-4090-9780]{Joseph Glaser}
\affiliation{Department of Physics and Astronomy, West Virginia University, P.O. Box 6315, Morgantown, WV 26506, USA}
\affiliation{Center for Gravitational Waves and Cosmology, West Virginia University, Chestnut Ridge Research Building, Morgantown, WV 26505, USA}

\author[0000-0003-1884-348X]{Deborah C. Good}
\affiliation{Department of Physics and Astronomy, University of Montana, 32 Campus Drive, Missoula, MT 59812}

\author[0000-0002-1146-0198]{Kayhan G\"{u}ltekin}
\affiliation{Department of Astronomy and Astrophysics, University of Michigan, Ann Arbor, MI 48109, USA}

\author[0000-0003-2742-3321]{Jeffrey S. Hazboun}
\affiliation{Department of Physics, Oregon State University, Corvallis, OR 97331, USA}

\author[0000-0003-1082-2342]{Ross J. Jennings}
\altaffiliation{NANOGrav Physics Frontiers Center Postdoctoral Fellow}
\affiliation{Department of Physics and Astronomy, West Virginia University, P.O. Box 6315, Morgantown, WV 26506, USA}
\affiliation{Center for Gravitational Waves and Cosmology, West Virginia University, Chestnut Ridge Research Building, Morgantown, WV 26505, USA}

\author[0000-0002-7445-8423]{Aaron D. Johnson}
\affiliation{Center for Gravitation, Cosmology and Astrophysics, Department of Physics, University of Wisconsin-Milwaukee,\\ P.O. Box 413, Milwaukee, WI 53201, USA}
\affiliation{Division of Physics, Mathematics, and Astronomy, California Institute of Technology, Pasadena, CA 91125, USA}

\author[0000-0001-6295-2881]{David L. Kaplan}
\affiliation{Center for Gravitation, Cosmology and Astrophysics, Department of Physics, University of Wisconsin-Milwaukee,\\ P.O. Box 413, Milwaukee, WI 53201, USA}

\author[0000-0002-6625-6450]{Luke Zoltan Kelley}
\affiliation{Department of Astronomy, University of California, Berkeley, 501 Campbell Hall \#3411, Berkeley, CA 94720, USA}

\author[0000-0003-0123-7600]{Joey S. Key}
\affiliation{University of Washington Bothell, 18115 Campus Way NE, Bothell, WA 98011, USA}

\author[0000-0002-9197-7604]{Nima Laal}
\affiliation{Department of Physics, Oregon State University, Corvallis, OR 97331, USA}

\author[0000-0003-0721-651X]{Michael T. Lam}
\affiliation{SETI Institute, 339 N Bernardo Ave Suite 200, Mountain View, CA 94043, USA}
\affiliation{School of Physics and Astronomy, Rochester Institute of Technology, Rochester, NY 14623, USA}
\affiliation{Laboratory for Multiwavelength Astrophysics, Rochester Institute of Technology, Rochester, NY 14623, USA}

\author[0000-0003-1096-4156]{William G. Lamb}
\affiliation{Department of Physics and Astronomy, Vanderbilt University, 2301 Vanderbilt Place, Nashville, TN 37235, USA}

\author[0000-0001-6436-8216]{Bjorn Larsen}
\affiliation{Department of Physics, Yale University, New Haven, CT 06520, USA}

\author{T. Joseph W. Lazio}
\affiliation{Jet Propulsion Laboratory, California Institute of Technology, 4800 Oak Grove Drive, Pasadena, CA 91109, USA}

\author[0000-0003-0771-6581]{Natalia Lewandowska}
\affiliation{Department of Physics and Astronomy, State University of New York at Oswego, Oswego, NY 13126, USA}

\author[0000-0001-5766-4287]{Tingting Liu}
\affiliation{Department of Physics and Astronomy, West Virginia University, P.O. Box 6315, Morgantown, WV 26506, USA}
\affiliation{Center for Gravitational Waves and Cosmology, West Virginia University, Chestnut Ridge Research Building, Morgantown, WV 26505, USA}

\author[0000-0001-5373-5914]{Jing Luo}
\altaffiliation{Deceased}
\affiliation{Department of Astronomy \& Astrophysics, University of Toronto, 50 Saint George Street, Toronto, ON M5S 3H4, Canada}

\author[0000-0001-5229-7430]{Ryan S. Lynch}
\affiliation{Green Bank Observatory, P.O. Box 2, Green Bank, WV 24944, USA}

\author[0000-0002-4430-102X]{Chung-Pei Ma}
\affiliation{Department of Astronomy, University of California, Berkeley, 501 Campbell Hall \#3411, Berkeley, CA 94720, USA}
\affiliation{Department of Physics, University of California, Berkeley, CA 94720, USA}

\author[0000-0003-2285-0404]{Dustin R. Madison}
\affiliation{Department of Physics, University of the Pacific, 3601 Pacific Avenue, Stockton, CA 95211, USA}

\author[0000-0001-5481-7559]{Alexander McEwen}
\affiliation{Center for Gravitation, Cosmology and Astrophysics, Department of Physics, University of Wisconsin-Milwaukee,\\ P.O. Box 413, Milwaukee, WI 53201, USA}

\author[0000-0002-2885-8485]{James W. McKee}
\affiliation{Department of Physics and Astronomy, Union College, Schenectady, NY 12308, USA}

\author[0000-0001-7697-7422]{Maura A. McLaughlin}
\affiliation{Department of Physics and Astronomy, West Virginia University, P.O. Box 6315, Morgantown, WV 26506, USA}
\affiliation{Center for Gravitational Waves and Cosmology, West Virginia University, Chestnut Ridge Research Building, Morgantown, WV 26505, USA}

\author[0000-0002-2689-0190]{Patrick M. Meyers}
\affiliation{Division of Physics, Mathematics, and Astronomy, California Institute of Technology, Pasadena, CA 91125, USA}

\author[0000-0002-4307-1322]{Chiara M. F. Mingarelli}
\affiliation{Department of Physics, Yale University, New Haven, CT 06520, USA}

\author[0000-0003-2898-5844]{Andrea Mitridate}
\affiliation{Deutsches Elektronen-Synchrotron DESY, Notkestr. 85, 22607 Hamburg, Germany}

\author[0009-0006-8387-2040]{Jonathan Nay}
\affiliation{Department of Physics, The University of Texas at Austin, Austin, TX 78712, USA}

\author[0000-0002-6709-2566]{David J. Nice}
\affiliation{Department of Physics, Lafayette College, Easton, PA 18042, USA}

\author[0000-0002-4941-5333]{Stella Koch Ocker}
\affiliation{Division of Physics, Mathematics, and Astronomy, California Institute of Technology, Pasadena, CA 91125, USA}
\affiliation{The Observatories of the Carnegie Institution for Science, Pasadena, CA 91101, USA}

\author[0000-0002-2027-3714]{Ken D. Olum}
\affiliation{Institute of Cosmology, Department of Physics and Astronomy, Tufts University, Medford, MA 02155, USA}

\author[0000-0001-5465-2889]{Timothy T. Pennucci}
\affiliation{Institute of Physics and Astronomy, E\"{o}tv\"{o}s Lor\'{a}nd University, P\'{a}zm\'{a}ny P. s. 1/A, 1117 Budapest, Hungary}

\author[0000-0001-5681-4319]{Polina Petrov}
\affiliation{Department of Physics and Astronomy, Vanderbilt University, 2301 Vanderbilt Place, Nashville, TN 37235, USA}

\author[0000-0002-8826-1285]{Nihan S. Pol}
\affiliation{Department of Physics, Texas Tech University, Box 41051, Lubbock, TX 79409, USA}

\author[0000-0002-2074-4360]{Henri A. Radovan}
\affiliation{Department of Physics, University of Puerto Rico, Mayag\"{u}ez, PR 00681, USA}

\author[0000-0001-5799-9714]{Scott M. Ransom}
\affiliation{National Radio Astronomy Observatory, 520 Edgemont Road, Charlottesville, VA 22903, USA}

\author[0000-0002-5297-5278]{Paul S. Ray}
\affiliation{Space Science Division, Naval Research Laboratory, Washington, DC 20375-5352, USA}


\author[0000-0001-8557-2822]{Jessie C. Runnoe}
\affiliation{Department of Physics and Astronomy, Vanderbilt University, 2301 Vanderbilt Place, Nashville, TN 37235, USA}

\author[0000-0001-7832-9066]{Alexander Saffer}
\altaffiliation{NANOGrav Physics Frontiers Center Postdoctoral Fellow}
\affiliation{National Radio Astronomy Observatory, 520 Edgemont Road, Charlottesville, VA 22903, USA}

\author[0009-0006-5476-3603]{Shashwat C. Sardesai}
\affiliation{Center for Gravitation, Cosmology and Astrophysics, Department of Physics, University of Wisconsin-Milwaukee,\\ P.O. Box 413, Milwaukee, WI 53201, USA}

\author[0000-0003-2807-6472]{Kai Schmitz}
\affiliation{Institute for Theoretical Physics, University of MÃ¼nster, 48149 MÃ¼nster, Germany}

\author[0000-0002-7778-2990]{Xavier Siemens}
\affiliation{Department of Physics, Oregon State University, Corvallis, OR 97331, USA}
\affiliation{Center for Gravitation, Cosmology and Astrophysics, Department of Physics, University of Wisconsin-Milwaukee,\\ P.O. Box 413, Milwaukee, WI 53201, USA}

\author[0000-0003-1407-6607]{Joseph Simon}
\altaffiliation{NSF Astronomy and Astrophysics Postdoctoral Fellow}
\affiliation{Department of Astrophysical and Planetary Sciences, University of Colorado, Boulder, CO 80309, USA}

\author[0000-0002-1530-9778]{Magdalena S. Siwek}
\affiliation{Center for Astrophysics, Harvard University, 60 Garden St, Cambridge, MA 02138, USA}

\author[0000-0003-2685-5405]{Tristan L. Smith}
\affiliation{Department of Physics and Astronomy, Swarthmore College, Swarthmore, PA 19081, USA}

\author[0000-0002-5176-2924]{Sophia V. Sosa Fiscella}
\affiliation{School of Physics and Astronomy, Rochester Institute of Technology, Rochester, NY 14623, USA}
\affiliation{Laboratory for Multiwavelength Astrophysics, Rochester Institute of Technology, Rochester, NY 14623, USA}

\author[0000-0001-9784-8670]{Ingrid H. Stairs}
\affiliation{Department of Physics and Astronomy, University of British Columbia, 6224 Agricultural Road, Vancouver, BC V6T 1Z1, Canada}

\author[0000-0002-1797-3277]{Daniel R. Stinebring}
\affiliation{Department of Physics and Astronomy, Oberlin College, Oberlin, OH 44074, USA}

\author[0000-0002-2820-0931]{Abhimanyu Susobhanan}
\affiliation{Max-Planck-Institut fÃ¼r Gravitationsphysik (Albert-Einstein-Institut), Callinstrasse 38, D-30167, Hannover, Germany}

\author[0000-0002-1075-3837]{Joseph K. Swiggum}
\altaffiliation{NANOGrav Physics Frontiers Center Postdoctoral Fellow}
\affiliation{Department of Physics, Lafayette College, Easton, PA 18042, USA}

\author{Jacob Taylor}
\affiliation{Department of Physics, Oregon State University, Corvallis, OR 97331, USA}

\author[0000-0003-0264-1453]{Stephen R. Taylor}
\affiliation{Department of Physics and Astronomy, Vanderbilt University, 2301 Vanderbilt Place, Nashville, TN 37235, USA}

\author[0000-0002-2451-7288]{Jacob E. Turner}
\affiliation{Green Bank Observatory, P.O. Box 2, Green Bank, WV 24944, USA}

\author[0000-0001-8800-0192]{Caner Unal}
\affiliation{Department of Physics, Middle East Technical University, 06531 Ankara, Turkey}
\affiliation{Department of Physics, Ben-Gurion University of the Negev, Be'er Sheva 84105, Israel}
\affiliation{Feza Gursey Institute, Bogazici University, Kandilli, 34684, Istanbul, Turkey}

\author[0000-0002-4162-0033]{Michele Vallisneri}
\affiliation{Jet Propulsion Laboratory, California Institute of Technology, 4800 Oak Grove Drive, Pasadena, CA 91109, USA}
\affiliation{Division of Physics, Mathematics, and Astronomy, California Institute of Technology, Pasadena, CA 91125, USA}

\author[0000-0002-6428-2620]{Rutger van~Haasteren}
\affiliation{Max-Planck-Institut fÃ¼r Gravitationsphysik (Albert-Einstein-Institut), Callinstrasse 38, D-30167, Hannover, Germany}

\author{Joris Verbiest}
\affiliation{Department of Physics, University of Central Florida, Orlando, FL 32816-2385, USA}

\author[0000-0003-4700-9072]{Sarah J. Vigeland}
\affiliation{Center for Gravitation, Cosmology and Astrophysics, Department of Physics, University of Wisconsin-Milwaukee,\\ P.O. Box 413, Milwaukee, WI 53201, USA}

\author[0000-0002-6020-9274]{Caitlin A. Witt}
\affiliation{Center for Interdisciplinary Exploration and Research in Astrophysics (CIERA), Northwestern University, Evanston, IL 60208, USA}
\affiliation{Adler Planetarium, 1300 S. DuSable Lake Shore Dr., Chicago, IL 60605, USA}

\author[0000-0003-1562-4679]{David Wright}
\affiliation{Department of Physics, Oregon State University, Corvallis, OR 97331, USA}

\author[0000-0002-0883-0688]{Olivia Young}
\affiliation{School of Physics and Astronomy, Rochester Institute of Technology, Rochester, NY 14623, USA}
\affiliation{Laboratory for Multiwavelength Astrophysics, Rochester Institute of Technology, Rochester, NY 14623, USA}

\collaboration{1000}{The NANOGrav Collaboration}
\noaffiliation
\correspondingauthor{Jonathan Nay}
\email{jonathan.nay@nanograv.org}

\begin{abstract}
Pulsar timing array observations have found evidence for an isotropic gravitational wave background with the Hellings-Downs angular correlations, expected from general relativity. This interpretation hinges on the measured shape of the angular correlations, which is predominately quadrupolar under general relativity. Here we explore a more flexible parameterization: we expand the angular correlations into a sum of Legendre polynomials and use a Bayesian analysis to constrain their coefficients with the 15-year pulsar timing data set collected by the North American Nanohertz Observatory for Gravitational Waves (NANOGrav). When including Legendre polynomials with multipoles $\ell \geq 2$, we only find a significant signal in the quadrupole with an amplitude consistent with general relativity and non-zero at the $\sim 95\%$ confidence level and a Bayes factor of 200. When we include multipoles $\ell \leq 1$, the Bayes factor evidence for quadrupole correlations decreases by more than an order of magnitude due to evidence for a monopolar signal at approximately 4~nHz which has also been noted in previous analyses of the NANOGrav 15-year data. Further work needs to be done in order to better characterize the properties of this monopolar signal and its effect on the evidence for quadrupolar angular correlations. 
\end{abstract}


\section{\label{sec:intro}Introduction} 

Several independent pulsar timing array (PTA) observations have found evidence for a gravitational wave background (GWB) in the nano-Hertz (nHz) frequency band with high levels of significance~\citep{15yrGWB, EPTA:2023fyk, Reardon:2023gzh, Xu:2023wog}. This GWB may have been produced by a population of unresolved supermassive black hole binaries (SMBHBs) \citep{NANOGrav:2023hfp}, exotic processes in the early universe that source a cosmological GWB~\citep{Boddy:2022knd, Caldwell:2022qsj, Green:2022hhj}, or a combination of both \citep{NG15new_physics}.

PTA observations measure pulses of radio emission from millisecond pulsars, which serve as precise astronomical clocks due to their highly stable rotational periods~\citep{Matsakis:1997,Hobbs:2012,Hobbs:2020a}. Gravitational waves (GWs) cause shifts in the pulse times of arrival (TOAs), and PTA observations  achieve sensitivity to the effects of $\sim $1-100 nHz GWs by cross-correlating TOAs between pairs of pulsars~\citep{Sazhin:1978, Detweiler:1979wn, Maggiore:2018sht, MingCC2022}. Furthermore, for an isotropic stochastic GWB, these cross-correlations are purely a function of the angular separation between the pairs of pulsars on the sky, and general relativity (GR) predicts that they should have a predominately quadrupolar angular correlation known as the Hellings-Downs (HD) curve~\citep{hd83}. 

The detection of a significant cross-correlation consistent with the HD curve is considered essential in order to claim the detection of a GWB [see, e.g., \citep{Allen:2023kib}]. Deviations from this expectation may be due to mundane systematic effects such as errors in the solar system ephemerides which create a time-dependent dipolar correlation \citep{Roebber:2019gha,2020ApJ...893..112V} or errors in the correction of the time at the telescope to a common inertial time causing a time-dependent monopolar correlation \citep{Hobbs:2012,Hobbs:2020a} [see also \citep{Tiburzi:2015kqa, 12.5yrGWB}]. In addition, measuring the angular power spectrum may help identify the presence of anisotropies in the GWB~\citep{Mingarelli:2013dsa, Taylor:2013esa, Gair:2014rwa, Hotinli:2019tpc, AliHaimoudEtAl:2020}. More exotic possibilities, like deviations from GR, also affect the detailed shape of the pulsar-pair angular correlations \citep{2008ApJ...685.1304L,Chamberlin:2011ev,Cornish:2017oic,NANOGrav:2021ini}. 

The North American Nanohertz Observatory for Gravitational Waves (NANOGrav) collaboration has used its 15-year pulsar timing data set to search for a GWB \citet[hereafter \citetalias{15yrGWB}]{15yrGWB}. The Bayesian analyses performed in \citetalias{15yrGWB} and by other PTA collaborations \citep{EPTA:2023fyk, Reardon:2023gzh, Xu:2023wog} focused on establishing the evidence for the HD cross-correlations over an analysis that neglects the cross-correlations altogether. Here we use a more flexible parameterization of the shape of the angular cross-correlations by expanding it into a sum of Legendre polynomials with free coefficients $c_{\ell}$; we refer to this as a ``harmonic analysis'' \citep{HAforPTAs}. GR predicts the angular power spectrum has a dominant quadrupole ($\ell=2$) contribution due to the two tensor polarization modes of GWs, while higher multipole contributions scale as $\sim$$\ell^{-3}$~\citep{Gair:2014rwa, Qin:2018yhy}.

Consistent with the \citetalias{15yrGWB} results and with the predictions of an isotropic GWB in GR, we find strong evidence \citep{jeffreys1998theory} (a Bayes factor of 200) for the dominant quadrupole correlations in the NANOGrav 15-year data with $c_2/c_2^{\rm HD} = 1.088^{+0.32}_{-0.45}$, and there is no evidence for multipoles higher than the quadrupole. When we include monopole correlations in our analyses, the quadrupole evidence is reduced by more than an order of magnitude due to the presence of a monopolar signal in the data at $\approx \!\! 4$ nHz. This monopolar signal has been extensively investigated \citep{15yrGWB,NG15new_physics} but currently has no clear explanation. 

Previous work on parameterizing the shape of the pulsar-pair cross-correlations has focused on using a minimum variance estimator [i.e., the optimal statistic \citep{OPTSTAT0,2013ApJ...762...94D,2015PhRvD..91d4048C,Vigeland:2018ipb}]. In particular, the multiple component optimal statistic (MCOS) \citep{Sardesai:2023qsw} allows for an estimate of the $c_{\ell}$s that broadly agrees with the harmonic analysis we present here. The differences that we find are likely due to the fact that the current MCOS approach does not properly take into account the full cross-correlation between pulsars. Furthermore, some MCOS analyses that have appeared in the literature (e.g., \citetalias{15yrGWB}) account only for the uncertainty in the estimator itself, leaving out the much larger contribution due to marginalizing over the uncertainty in the GWB amplitude. Most importantly, the Bayesian analysis we present here directly utilizes the PTA likelihood when exploring the inferred shape of the angular cross-correlations expanded in Legendre polynomials. 

The paper is organized as follows. In~\S\ref{sec:methods} we provide background for our harmonic analysis approach, discuss our modeling methodologies, and list the various models we use in this paper. In~\S\ref{sec:analyses} we provide the results of our GWB harmonic analyses on previous NANOGrav data sets, investigate alternative monopole and dipole correlations, and examine frequency-dependent angular-correlation models. In~\S\ref{sec:previous_work} we compare our results to previous work parameterizing the shape of the pulsar pair angular correlation.  We discuss our results and summarize our conclusions in~\S\ref{sec:discussion}.

\section{\label{sec:methods}Harmonic Analysis Methods}

The Bayesian analysis of the NANOGrav 15y data is identical to what is done in \citetalias{15yrGWB}; in particular, the likelihood function is given by~\citep{NANOGrav:2023icp}
\begin{equation}
  p(\vec{\delta t} \: | \: \vec{\eta}) = \frac{1}{\sqrt{\text{det}(2 \pi \mathcal{C})}} \: \text{exp} \left(-\frac{1}{2}\vec{\delta r}^T \mathcal{C}^{-1} \vec{\delta r} \right),
  \label{eq:Likelihood}
\end{equation}
where $\vec{\delta t}$ are the pulsar timing residuals, $\vec{\eta}$ are the model parameters, and $\mathcal{C}$ is a covariance matrix. The covariance matrix consists of several sources of white noise for each pulsar whose parameters are set to the maximum likelihood of an analyses of individual pulsars \citep{NANOGrav:2023ctt}. The residual vector is defined to be 
\begin{equation}
  \vec{\delta r} \equiv \vec{\delta t} - \mathcal{F} \vec c - \mathcal{M} \vec \epsilon
\end{equation}
where $\mathcal{F} \vec c$ is a Gaussian process that models intrinsic and correlated red-noise processes and $\mathcal{M} \vec \epsilon$ takes into account variations in the deterministic timing model for each pulsar. The set of $\vec c$ is drawn from a zero-mean Gaussian with covariance 
\begin{equation}
    \langle c_{ai}c_{bj}\rangle = \delta_{ij} \left(\delta_{ab} \varphi_{a,i}+S_{ab,i}\right),\label{eq:c_cov}
\end{equation}
where $a,b$ range over pulsars and $i,j$ over Fourier components, which are then transformed into the time domain. The term $\varphi_{a,i}$ models the spectrum of intrinsic red noise in pulsar $a$, which is modeled as a power law,
\begin{equation}
  P_{\text{RN},a}(f) \equiv \frac{A^{2}_{\text{RN},a}}{12 \pi^{2}} \: \left(\frac{f}{f_{\text{yr}}}\right)^{-\gamma_{\text{RN},a}}\: f_{\text{yr}}^{-3},
  \label{eq:RNpowerlaw}
\end{equation}
where $A_{\text{RN},a}$ is the dimensionless amplitude for the intrinsic red noise of pulsar $a$, and $\gamma_{\text{RN},a}$ is the corresponding spectral index. The term $S_{ab,i}$ models a stochastic process that is correlated across all pulsars, and the auto-correlation $S_{aa,i}$ is the same for all pulsars.  We refer the reader to~\citep{NANOGrav:2023icp} for a more detailed discussion of the likelihood.

The correlated stochastic process can be expressed in the general form
\begin{equation}
  S_{ab}(f) = P(f) \: \Gamma_{ab}(f),
  \label{eq:genericcrossPSD}
\end{equation}
where $P(f)$ is the frequency power spectrum, and $\Gamma_{ab}(f)$ is the angular correlation function (which may, in general, depend on frequency).
We follow the convention of referring to terms in our model as a cross-correlation when $a$ and $b$ denote distinct pulsars, and an auto-correlation when $a$ and $b$ denote the same pulsar. 

We describe the specific angular-correlation parameterizations used in this paper in~\S\ref{subsec:methods_angular}.
We describe our frequency power spectrum models in~\S\ref{subsec:methods_frequency}.
The Bayesian analysis models used in this paper are provided in~\S\ref{subsec:methods_models}.
Our method of calculating model evidence and individual angular-correlation evidence is described in~\S\ref{subsec:methods_evidence}.

\subsection{\label{subsec:methods_angular}GWB Angular-Correlation Models}

GR predicts that an isotropic stochastic GWB induces a frequency-independent angular correlation between pulsar pairs given by the HD curve~\citep{hd83}
\begin{align}
\begin{split}
  \Gamma^{\rm HD}_{ab} & =  (1 + \delta_{ab}) \: \Bigg[ \frac{1}{2}  - \frac{1}{4} \left(\frac{1-\cos{\theta_{ab}}}{2} \right) \\
  & + \frac{3}{2} \left(\frac{1-\cos{\theta_{ab}}}{2} \right) \: \log{\left(\frac{1-\cos{\theta_{ab}}}{2} \right)} \Bigg]
  \label{eq:HDcurve}
\end{split}
\end{align}
where $\cos{\theta_{ab}} = \hat{n}_a \cdot \hat{n}_b$ for pulsars $a$ and $b$ located on the sky at $\hat{n}_a$ and $\hat{n}_b$, respectively, and $\delta_{ab}$ comes from the pulsar term, which is relevant for co-located pulsars~\citep{hd83, Anholm:2008wy, Mingarelli:2013dsa}.

An equivalent representation of the HD curve using a Legendre polynomial expansion gives the angular-correlation function~\citep{Gair:2014rwa, Roebber:2016jzl}
\begin{equation}
  \Gamma^{\rm HD}_{ab} = (1 + \delta_{ab}) \: \sum_{\ell = 2}^{\infty} c^{\rm HD}_\ell P_\ell(\cos\theta_{ab}),
  \label{eq:LegendreSumGR}
\end{equation}
where $P_\ell$ are Legendre polynomials with coefficients
\begin{equation}
  c^{\rm HD}_{\ell} = \frac{3}{2} \: (2\ell+1) \: \frac{(\ell-2)!}{(\ell+2)!}
  \label{eq:LegendreCoefficientsGR}
\end{equation}
for $\ell \ge 2$ and $c^{\rm HD}_0 = c^{\rm HD}_1 = 0$.
The HD Legendre coefficients exhibit a dominant quadrupolar contribution and a sharp drop off ($\propto \ell^{-3}$) at higher multipoles.

In our harmonic analysis, we follow \cite{HAforPTAs} and parameterize the angular correlations by
\begin{align}
  \Gamma^{\ell_{\rm max}}_{ab} & = (1-\delta_{ab}) \sum_{\ell=2}^{\ell_{\rm max}} c_\ell P_\ell(\cos\theta_{ab}) + \delta_{ab}.
  \label{eq:LegendreSum}
\end{align}
We note that, unlike the HD curve in \autoref{eq:HDcurve}, this parameterization separates the auto- and cross-correlations into two distinct terms. There are both benefits and drawbacks to this approach. 

The parameterization in \autoref{eq:LegendreSum} allows us to directly test for the presence of cross-correlations, since in the limit that $c_\ell = 0$, we are left with only a contribution to the auto-correlation. As a result, a non-zero detection of any $c_\ell$ is evidence for a non-zero cross-correlation. In addition, the inferred shape of the angular correlations comes solely from the pulsar cross-correlations, which are less affected by processes that are intrinsic to each pulsar. Finally, this choice also allows us to directly compare the constraints found here to previous methods used to characterize the shape of the angular-correlations, which also exclusively rely on the pulsar cross-correlations (see ~\S\ref{sec:previous_work}). 

On the other hand, specific models that predict modifications to the pulsar-pair angular correlations have a particular relationship between the $c_\ell$'s and the auto-correlation, $\Gamma_{aa}$. Many models (including GR and those that predict sub-luminal propagation of tensor GWs, such as massive gravity) have $\Gamma_{aa} = 2 \sum_{\ell=2}^\infty c_\ell$. Other models, such as those that predict the existence of a scalar longitudinal polarization, have pulsar distance dependent auto-correlations [see, e.g., \citep{2008ApJ...685.1304L}]. Therefore, estimates of the $c_\ell$'s using the parameterization in \autoref{eq:LegendreSum} cannot generally be used to directly constrain specific theories. 

Finally, we note that the parameterization in \autoref{eq:LegendreSum} only contains multipoles with $\ell \geq 2$. This restriction is motivated by the expectation that the angular correlations described by these multipoles all share the same frequency power spectrum, $P(f)$ (see \autoref{eq:genericcrossPSD}). Lower multipoles may be excited by non-standard GW polarizations, unmodeled effects on the timing of the pulsars, and shifts in the solar system barycenter. Such effects, in general, come with a different dependence on time. Given this, we model these lower multipoles using a separate frequency power spectrum, as discussed in more detail in the next subsection and in Table \ref{tab:table_models}. 

\renewcommand{\arraystretch}{1.3}
\begin{table*}[t]
  \centering
  \begin{tabular}{|p{3cm}|p{6cm}|p{3.5cm}|p{3.5cm}|}
    \hline
    \centering Model Name & \centering Model Description & \centering Angular-correlation parameterization & \centering Frequency Power-spectrum parameterization \tabularnewline
    \hline
    $\text{HA}^\gamma(c_{2},\dots,c_{\ell_{\rm max}})$ & Parameterized angular correlations & $\Gamma^{\ell_{\rm max}}_{ab}$ (\autoref{eq:LegendreSum}) & $P_{\text{gw}}(f)$ (\autoref{eq:GWpowerlaw}) \\
    \hline
    $\text{HD}^{\gamma}$ & Fixed HD angular correlations  & $\Gamma^{\rm HD}_{ab}$ (\autoref{eq:LegendreSumGR})  & $P_{\text{gw}}(f)$  \\
    \hline
    $\text{CURN}^{\gamma}$ & Common uncorrelated red noise & $\Gamma^{\rm CURN}_{ab} =\delta_{ab}$ & $P_{\text{gw}}(f)$  \\
    \hline
    $\text{IRN}$ & Pulsar intrinsic red noise & $\Gamma^{\rm IRN}_{ab} = \delta_{ab}$ & $P_{\text{RN}, a}(f)$ (\autoref{eq:RNpowerlaw}) \\
    \hline
    $\text{MONO}^{\rm free}$ & Monopole free spectrum & $\Gamma^{\rm MONO}_{ab} = 1$ & $\Phi^2(f) \: T_{\text{obs}}$ (\autoref{eq:FreeSpectrum}) \\
    \hline
    $\text{DIP}^{\rm free}$ & Dipole free spectrum & $\Gamma^{\rm DIP}_{ab} = \cos \theta_{ab}$ & $\Phi^2(f) \: T_{\text{obs}}$ \\
    \hline
  \end{tabular}
  \caption{List of models used in our analyses.
  Each model is the product of the angular-correlation parameterization and the frequency power spectrum parameterization listed above.
  The GWB harmonic analysis model in the first row, $\text{HA}^\gamma(c_{\ell})$, comes from~\cite{HAforPTAs} and is the primary model for this paper.
  The remaining models listed above follow the same naming convention as in \citetalias{15yrGWB}.}
  \label{tab:table_models}
\end{table*}
\renewcommand{\arraystretch}{1}

\subsection{\label{subsec:methods_frequency}Frequency Power-Spectrum Models}

We model the frequency power spectrum for the GWB as a power law
\begin{equation}
  P_{\text{gw}}(f) \equiv \frac{A^{2}_{\text{gw}}}{12 \pi^{2}} \: \left(\frac{f}{f_{\text{yr}}}\right)^{-\gamma_{\text{gw}}}\: f_{\text{yr}}^{-3},
  \label{eq:GWpowerlaw}
\end{equation}
where $A_{\text{gw}}$ is the dimensionless strain amplitude of the GWB at a reference frequency $f_{\text{yr}}=1/\text{year}$, and $\gamma_{\text{gw}}$ is the spectral index.
We expect $\gamma_{\text{gw}} \simeq 13/3$ for a source of inspiraling SMBHBs~\citep{Phinney:2001di}.

There is a wide range of possibilities for the frequency spectra associated with monopole and dipole correlations. Therefore, the frequency power-spectrum for a monopole and dipole is modeled as an independent parameter for each GWB frequency.
This approach is referred to as a ``free-spectrum'' model (e.g., see \citetalias{15yrGWB}).
Using the same approach as in \citetalias{15yrGWB}, we define the free-spectrum parameter for the $i^{th}$ frequency 
\begin{equation}
  \Phi^2(f_i) \equiv P(f_i) \Delta f = P(f_i) / T_{\text{obs}},
  \label{eq:FreeSpectrum}
\end{equation}
where $\Delta f$ is the frequency resolution, which is set by $T_{\text{obs}}$, the longest observational baseline of all pulsars in the data set.

\subsection{\label{subsec:methods_models}Bayesian Analysis Models}

The baseline GWB harmonic analysis is the product of \autoref{eq:LegendreSum} and \autoref{eq:GWpowerlaw} which gives
\begin{equation}
  S_{ab}(f) = \frac{A^{2}_{\text{gw}}}{12 \pi^{2}} \: \left(\frac{f}{f_{\text{yr}}}\right)^{-\gamma_{\text{gw}}}\: f_{\text{yr}}^{-3} \: \Gamma^{\ell_{\rm max}}_{ab}.
  \label{eq:modelGWB}
\end{equation}
We denote this model by $\text{HA}^\gamma(c_{2},\dots,c_{\ell_{\rm max}})$.

\autoref{tab:table_models} provides the models used in this paper, which are similarly obtained by multiplying an angular-correlation model with a frequency power-spectrum model.
The remaining models in \autoref{tab:table_models} are the same as those used use in \citetalias{15yrGWB}, and we use the same naming convention as \citetalias{15yrGWB}.
In particular, for the CURN model, which assumes no angular correlations, and the HD model, we use a $\gamma$ superscript to denote $\gamma_{\text{gw}}$ is a model parameter.
The pulsar intrinsic red noise (IRN) model is included for every pulsar in each analysis.

\subsection{\label{subsec:methods_evidence}Methods of Determining Evidence}

We compute the model evidence by comparing the Bayesian evidence for two different models using product-space sampling to determine the posterior odds ratio, as described in \citetalias{15yrGWB} and~\cite{NANOGrav:2023icp}.
For this paper, all Bayes factors are calculated by performing a model comparison with either a $\text{CURN}^{\gamma}$ model or an $\text{HD}^{\gamma}$ model.

We use the Savage-Dickey Bayes factor method~\citep{10.2307/2958475} to determine evidence for including a single multipole in our model.
The Savage-Dickey Bayes factor for a model with a single multiple $\ell$ is~\citep{HAforPTAs}
\begin{equation}
 \text{SD}_{\ell} = \frac{1}{p(c_{\ell} = 0)},
  \label{eq:SDequation}
\end{equation}
where $p(c_{\ell} = 0)$ is the probability of the Legendre coefficient's marginalized 1D posterior distribution evaluated at zero.
As discussed in Appendix B of~\cite{HAforPTAs}, this approach is justified because the lower bound of the prior range for each $c_{\ell}$ is zero (see~\S\Ref{subsec:analyses_inputs}), and setting the Legendre coefficient to zero removes the parameter from the model, as seen in \autoref{eq:LegendreSum}.

\section{\label{sec:analyses}Analyses and Results}

We use \texttt{ENTERPRISE}~\citep{ellis_justin_a_2020_4059815} and \texttt{enterprise-extensions}~\citep{enterprise-extensions} to calculate the likelihood function in \autoref{eq:Likelihood}.
We modify \texttt{enterprise-extensions} to include the Legendre coefficients as model parameters, as discussed in~\S\ref{subsec:methods_angular}, which is the same modification used in~\cite{HAforPTAs}.
We make additional modifications to \texttt{enterprise-extensions} to include free spectra models and frequency-dependent Legendre coefficient models discussed in~\S\Ref{subsec:methods_models}.
We also use the \texttt{HyperModel} package (referred to hereafter as hypermodel) of \texttt{enterprise-extensions} to calculate Bayes factors between pairs of models.
We use \texttt{PTMCMCSampler}~\citep{justin_ellis_2017_1037579} to perform Markov chain Monte Carlo (MCMC) sampling to determine parameter posterior distributions.

In~\S\ref{subsec:analyses_inputs} we describe the data sets, the white noise modeling technique, and the parameters used in our harmonic analyses.
In~\S\ref{subsec:analyses_evidence} we calculate the evidence for angular correlations in the NANOGrav data.
In~\S\ref{subsec:analyses_measurements} we analyze the multipole posterior distributions and use our results to reconstruct the angular-correlation function.
In~\S\ref{subsec:analyses_monopole} we extend our harmonic analyses to include monopole and dipole angular correlations.

\subsection{\label{subsec:analyses_inputs}Analysis Inputs and Parameters}

We analyze the \citetalias{15yrGWB} data set for the majority of the work presented in this paper. For comparison, we also analyze the NANOGrav 12.5-year data set analyzed in \citet[hereafter \citetalias{12.5yrGWB}]{12.5yrGWB}. We use the same pulsars in our harmonic analyses as in the original NANOGrav GWB papers; specifically, we only include pulsars with an observation time span greater than 3~years, which provides 67 and 45 pulsars for the \citetalias{15yrGWB} and \citetalias{12.5yrGWB} data sets, respectively.

For both the \citetalias{12.5yrGWB} and \citetalias{15yrGWB} data sets, we use the same number of frequencies ($N_f = 14$) for the GWB power spectrum as in their original analyses, with binned frequencies $f_i=i/T_{\rm obs}$ for $i=1,\dots,N_f$. For all analyses, including single pulsar noise modeling, we use $N_f=30$ for the pulsar intrinsic red noise power spectrum.
For monopole and dipole free-spectrum models, we use $N_f=5$, which covers the frequencies for which the evidence for a GWB is the strongest (as demonstrated in \citetalias{15yrGWB}).

The MCMC priors are given in \autoref{tab:table_inputs}.
For the Legendre coefficients, the lower end of the prior range comes from the requirement that the angular power spectrum is strictly positive, as shown in \autoref{eq:LegendreCoefficientsGR}, while the upper end comes from the requirement that the full pulsar covariance be positive definite.

\begin{table}[t]
  \centering
  \begin{tabular}{|p{1.7cm}|p{3.1cm}|p{2.7cm}|}
    \hline
    \centering MCMC & \centering MCMC & \centering Bayesian analysis\tabularnewline
    \centering parameter & \centering prior range & \centering model \tabularnewline
    \hline
    $\log_{10} A_{\text{gw}}$ & $U[-18,-11]$ & GWB Power-law\\
    $\gamma_{\text{gw}}$ & $U[0,7]$ & GWB Power-law \\
    $c_{\ell}$ & $U[0,1]$ & Harmonic Analysis\\
    \hline
    $\log_{10}\Phi(f_i)$ & $U[-9,-4]$ & Free-spectrum\\
    \hline
    $\log_{10} A_{\text{RN},a}$ & $U[-20,-11]$ & Pulsar IRN\\
    $\gamma_{\text{RN},a}$ & $U[0,7]$ & Pulsar IRN\\
    \hline
  \end{tabular}
  \caption{Prior ranges for the MCMC parameters. We denote uniform ranges by $U[x_{\rm min}, x_{\rm max}]$.}
  \label{tab:table_inputs}
    \vspace{-1\baselineskip}
\end{table}

We note that in addition to the priors listed in \autoref{tab:table_inputs}, there is an implicit prior on the $c_\ell$s due to their effect on the positive definiteness of the red-process covariance matrix given in \autoref{eq:c_cov}. If $\sum_{\ell =2}^{\ell_{\rm max}} c_{\ell} \gtrsim 1$, then the cross-correlations are larger than the GWB's contribution to the auto-correlations for pulsar-pair separation angles near $\theta_{ab}=0$. This can result in the red-process a covariance matrix in \autoref{eq:c_cov} that is not positive definite. The Cholesky decomposition algorithm used by \texttt{ENTERPRISE} to find the inverse of this covariance matrix then fails for these parameter values. This constraint imposes a prior $\sum_{\ell =2}^{\ell_{\rm max}} c_{\ell} \le 1$ so that the full prior on each Legendre coefficient is
\begin{equation}
  p(x) = N_\ell (1-x)^{N_\ell-1},
  \label{eq:Legendreprior}
\end{equation}
where $N_\ell$ is the total number of multipoles in the GWB model, and $x \in [0,1]$.

We run multiple MCMC chains in parallel to reduce processing time; we do not employ parallel tempering.
We combine sampling chains after removing a 25\% burn-in to create a single final chain. We use the Gelman-Rubin $R$-statistic~\cite{Gelman:1992zz} as a measure of chain convergence and require $R-1<0.05$ for all GWB parameters.

\subsection{\label{subsec:analyses_evidence} Bayesian Evidences}

For the \citetalias{15yrGWB} data set, we find that the Bayes factor of an $\text{HA}(c_2) / \text{HD}^{\gamma}$ hypermodel is $\approx \!\! 1$, which implies there is no preference for a model with quadrupole-only correlations over a model with HD correlations. \autoref{tab:table_analyses} provides the Bayes factors and information on the marginalized 1D posterior distributions for $c_2$ for various harmonic models, relative to a $\text{CURN}^{\gamma}$ model. Notably, we find Bayes factor of a $\text{HA}^\gamma(c_2) / \text{CURN}^{\gamma}$ hypermodel is $\approx \!\! 200$, consistent with the Bayes factor of $\approx \!\! 200$ for a $\text{HD}^{\gamma} / \text{CURN}^{\gamma}$ hypermodel for 14 GWB frequency components found in \citetalias{15yrGWB}.

\begin{table}[t]
\renewcommand{\arraystretch}{1}
  \centering
  \begin{tabular}{|p{2.26cm}|p{0.9cm}|p{0.6cm}|p{1.5cm}|p{1.5cm}|}
    \hline
    \centering Bayesian & \centering Bayes & \multicolumn{3}{c|}{Quadrupole ($c_2$)} \\
    \cline{3-5}
    \centering Model Name & \centering Factor & \centering Mean &  \centering 68\% CL &  \centering 95\% CL \tabularnewline
    \hline
     $\text{HA}^\gamma(c_2)$ & 200 & 0.34 & $[0.20, 0.44]$ & $[0.11, 0.58]$ \\
    \hline
     $\text{HA}^\gamma(c_2,c_3)$ & 55 & 0.30 & $[0.16, 0.41]$ & $[0.07, 0.56]$ \\
    \hline
     $\text{HA}^\gamma(c_2,c_3,c_4)$ & 0.9 & 0.28 & $[0.14, 0.38]$ & $[0.05, 0.52]$ \\
    \hline
     $\text{HA}^\gamma(c_2,...,c_5)$ & 0.1 & 0.27& $[0.14, 0.36]$ & $[0.05, 0.49]$ \\
    \hline
     $\text{HA}^\gamma(c_2)(45 \text{psrs})$ & 170 & 0.33 & $[0.19, 0.44]$ & $[0.10, 0.58]$ \\
    \hline
  \end{tabular}
\renewcommand{\arraystretch}{1}
  \caption{Summary of Bayes factors and quadrupole statistics from harmonic analyses of the \citetalias{15yrGWB} data set.
  The Bayes factors are calculated relative to a $\text{CURN}^{\gamma}$ model, which does not include angular correlations.
  The mean value of the quadrupole's marginalized 1D posterior distribution is provided at the 68\% and 95\% confidence levels (CLs).
  The analysis in the last row uses the \citetalias{15yrGWB} data set, but only includes the 45 pulsars in the \citetalias{12.5yrGWB} data set.}
  \label{tab:table_analyses}
\end{table}

When we include multipoles higher than the quadrupole, the evidence is reduced. For $\ell_{\rm max}$=3, 4, and~5, the $\text{HA}^\gamma(c_2,...,c_{\ell_{\rm max}}) / \text{CURN}^{\gamma}$ hypermodels give Bayes factors of approximately 55, 1, and 0.1, respectively.
These results are consistent with no evidence for multipoles $\ell \geq 4$ in our model. The Bayes factor of 55 for an $\text{HA}(c_2,c_3) / \text{CURN}^{\gamma}$ hypermodel suggests some evidence for the octupole may be present in the data; however, a model with quadrupole only correlations is highly preferred over a model with quadrupole and octupole correlations.

For the \citetalias{12.5yrGWB} data set, the Bayes factor of $\text{HA}^\gamma(c_2) / \text{CURN}^{\gamma}$ is $\approx \!\! 4$, consistent with the Bayes factors reported in \citetalias{12.5yrGWB} for $\text{HD}^{\gamma} / \text{CURN}^{\gamma}$.
To understand the reason for the large jump in quadrupole evidence going from the \citetalias{12.5yrGWB} data set to the \citetalias{15yrGWB} data set, we perform an additional analysis in which we use the \citetalias{15yrGWB} data set restricted to the 45 pulsars from the \citetalias{12.5yrGWB} data set, denoted $\text{HA}^\gamma(c_2)(45 \text{ psrs})$.
The Bayes factor of $\text{HA}^\gamma(c_2)(45 \text{ psrs})/ \text{CURN}^{\gamma}$ is 170, which is nearly the same as $\text{HA}^\gamma(c_2) / \text{CURN}^{\gamma}$. Thus, the large change in quadrupole evidence from the \citetalias{12.5yrGWB} data set to the \citetalias{15yrGWB} data set is primarily due to increasing the observation time span of the longest observed pulsars. This result is not surprising, because the 22 pulsars added between the \citetalias{12.5yrGWB} and the \citetalias{15yrGWB} data sets do not have long observation time spans, and therefore contribute less to lower frequencies where the GWB signal is expected to be the strongest. 

\begin{figure}[t]
  \centering
  \includegraphics[width=0.48\textwidth]{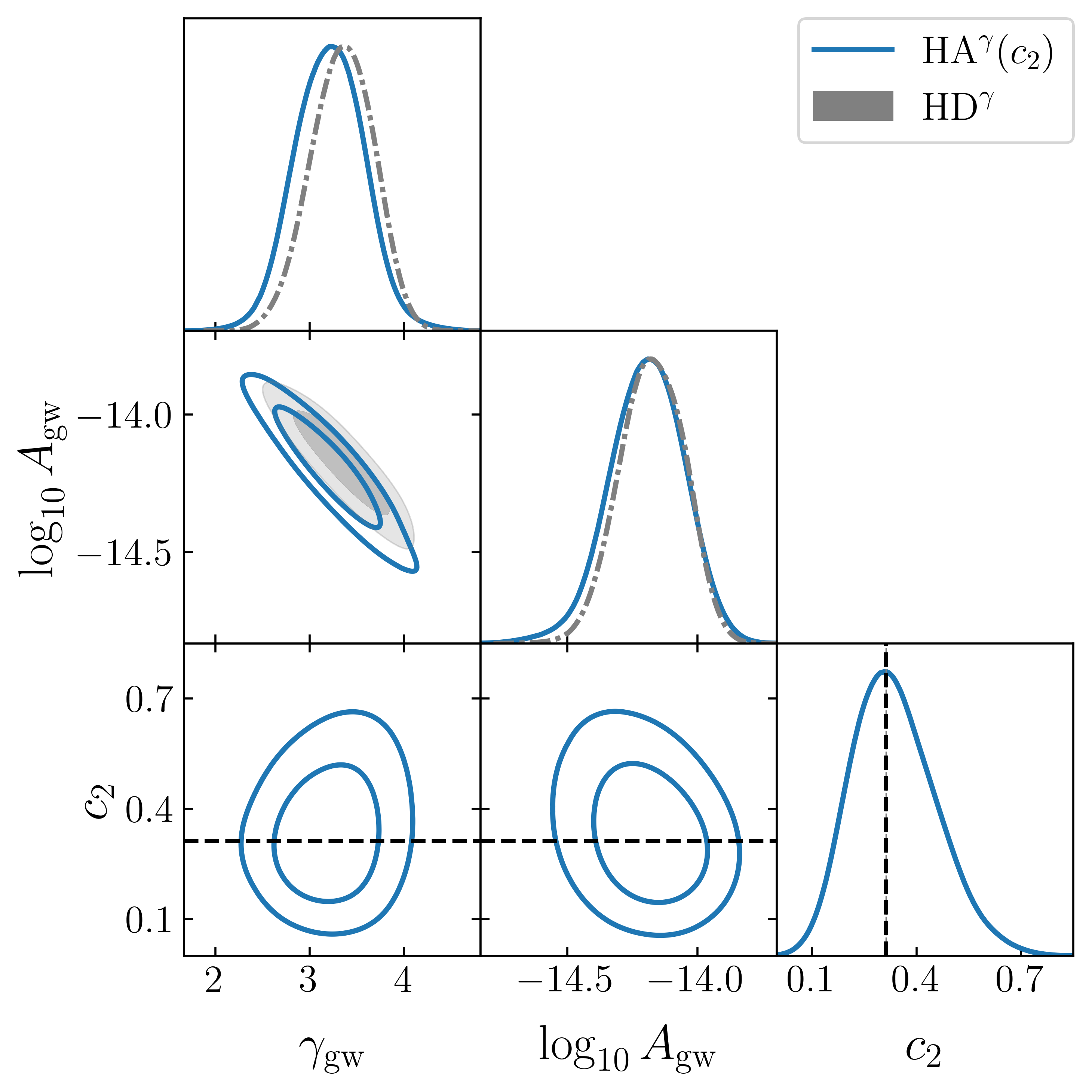}
  \caption{Marginalized 1D and 2D posterior distributions for quadrupole-only GWB harmonic analysis of the \citetalias{15yrGWB} data set for the model $\text{HA}^\gamma(c_2)$.
  The HD value for the quadrupole coefficient, $c_2^{\rm HD}=0.3125$ from \autoref{eq:LegendreCoefficientsGR}, is shown as the black dashed line.
  The GWB amplitude and spectral index from model $\text{HD}^{\gamma}$, which has angular correlations fixed to the theoretical HD values, are shown in gray.
  }
  \label{fig:CornerHA}
\end{figure}

\subsection{\label{subsec:analyses_measurements}Posterior Distributions}

We show the marginalized 1D and 2D posterior distributions of the three GWB parameters from model $\text{HA}^\gamma(c_2)$ when fit to the \citetalias{15yrGWB} data set in \autoref{fig:CornerHA}.
The posterior distribution of $c_2$ is consistent with the theoretical HD value of the quadrupole correlation, $c_2^{\rm HD}=0.3125$ from \autoref{eq:LegendreCoefficientsGR}, denoted by the dashed line in \autoref{fig:CornerHA}. The GWB amplitude and spectral index for an $\text{HD}^{\gamma}$ model, which has angular correlations fixed to the theoretical HD values, is shown in gray. We can see that the posterior distribution for $c_2$ is negligibly correlated with $\gamma_{\rm gw}$ and $\log_{10} A_{\rm gw}$.

\begin{figure*}[t]
  \centering
    \includegraphics[width =0.85\columnwidth]{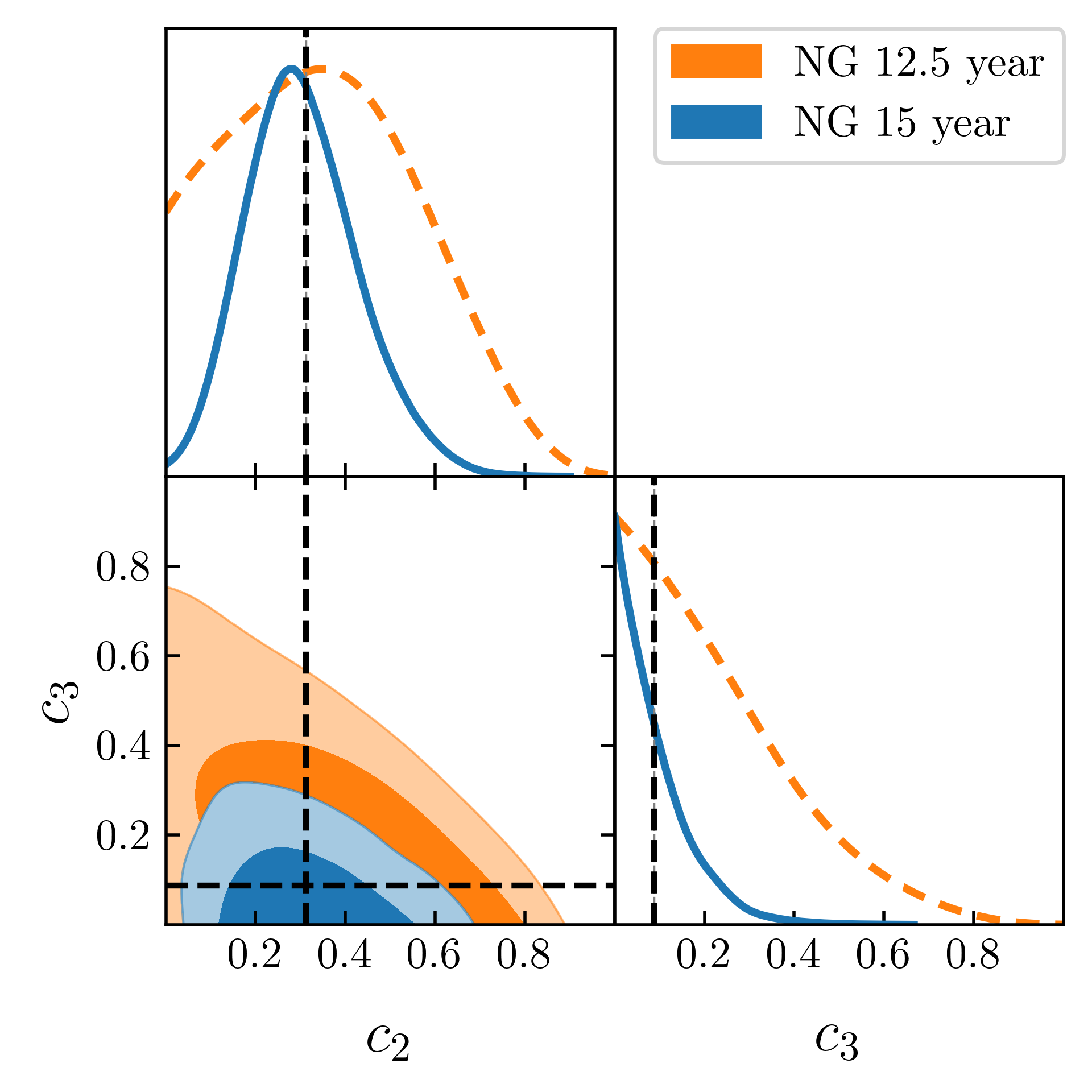}
        \includegraphics[width =1.15\columnwidth]{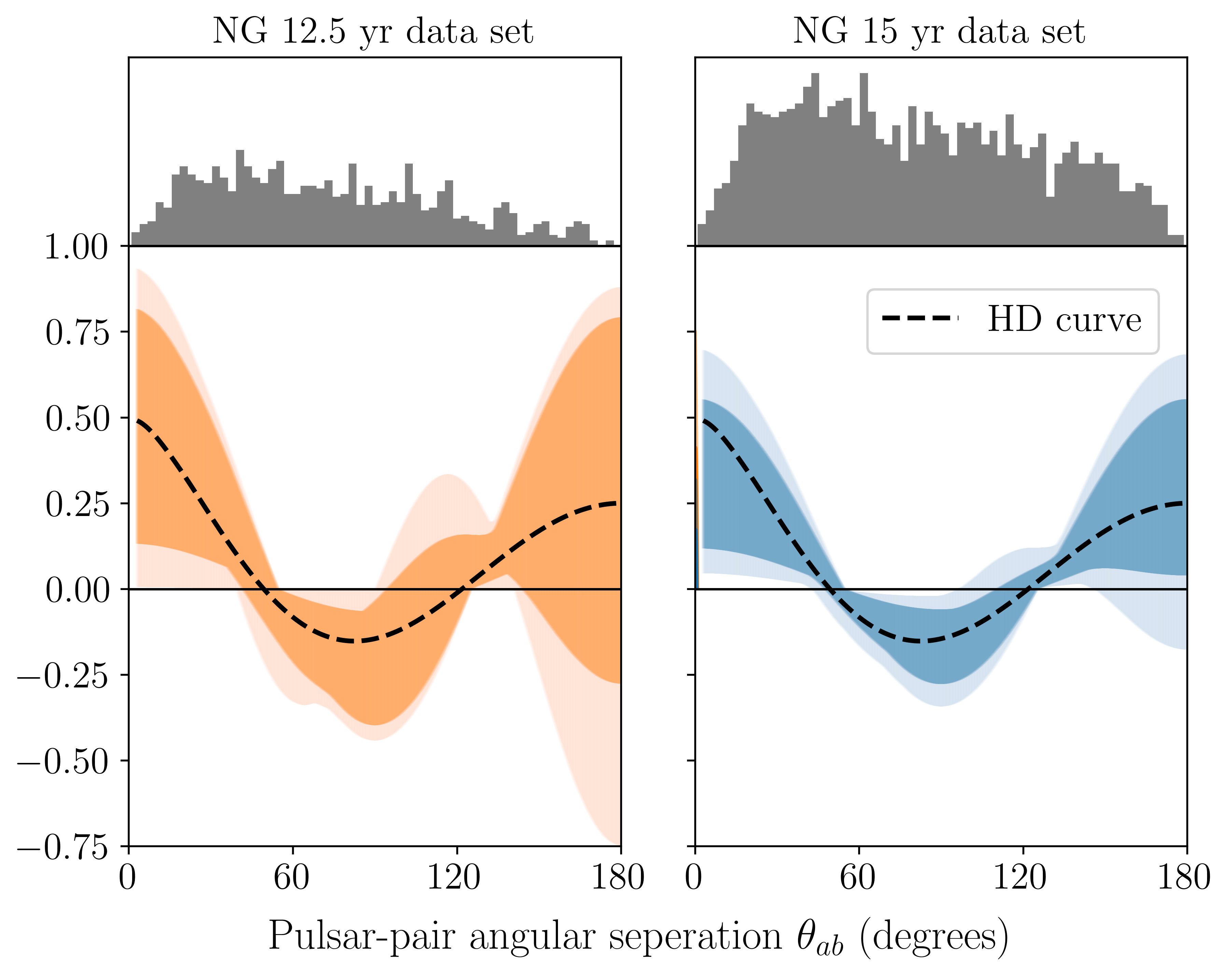}
  \caption{Left panel: Marginalized 1D and 2D posterior distributions of $c_2$ and $c_3$ for the harmonic analysis $\text{HA}^\gamma(c_2,c_3)$ of the NANOGrav 12.5 and 15 yr data sets.
  The dashed black lines show the HD value of each Legendre coefficient.
  Right panel: Reconstructed angular correlation function from the same $\text{HA}^\gamma(c_2,c_3)$ model with the NANOGrav 12.5 and 15 yr data sets.
  The dark- and light-shaded regions denote the 68\% and 95\% CL regions, respectively, from the marginalized 2D posterior distributions of the quadrupole and octupole parameters shown in the left panel.
  The dashed-black line is the HD curve, obtained from \autoref{eq:LegendreSumGR}.
  At the top of each plot, we provide a histogram of the pulsar-pair angular separations for these two NANOGrav data sets.
  }
    \label{fig:Csums}
\end{figure*}

The quadrupole's posterior distribution broadens as the number of Legendre coefficients in the model increases, as evident from \autoref{tab:table_analyses}. However, even for $\ell_{\rm max} = 5$, the posterior distribution of $c_2$ is consistent with $c_2^{\rm HD} = 0.3125$ and is non-zero at the 95\% CL. 

Using \autoref{eq:LegendreSum}, we can reconstruct the angular correlation function from the 68\% and 95\% contour regions of the marginalized 2D posterior distributions for $c_2$ and $c_3$.
The right panel of \autoref{fig:Csums} shows these reconstructions from the harmonic analyses of the \citetalias{12.5yrGWB} and \citetalias{15yrGWB} data sets with model $\text{HA}^\gamma(c_2,c_3)$. The dark- and light-shaded regions denote the 68\% CL and 95\% CL regions, respectively. The dashed black line is the HD curve from \autoref{eq:HDcurve}.
At the top of each reconstructed angular correlation function, we provide a histogram of pulsar-pair angular separations for the two data sets.

For both the \citetalias{12.5yrGWB} and \citetalias{15yrGWB} data sets, the HD curve lies within the 68\% contour region.
We see a large reduction in the spread of the reconstructed angular-correlation function going from the \citetalias{12.5yrGWB} data set to the \citetalias{15yrGWB} data set, as expected given the change in quadrupole evidence between these data sets.

Figure~4 of \cite{HAforPTAs} provides approximate scaling relationships for the mean-to-standard deviation ratio of the Legendre coefficient's marginalized 1D posterior distribution, as a function of the observation time and number of observed pulsars. For the quadrupole, this ratio is $\approx \!\! 2.5$ in the \citetalias{15yrGWB} data set, which is near the minimum value where the scaling relationships from \cite{HAforPTAs} begin to apply. For multipoles $\ell \geq 3$, this ratio is $\approx 1$ in all analyses.

\begin{figure}[t]
    \centering
    \includegraphics[width =\columnwidth]{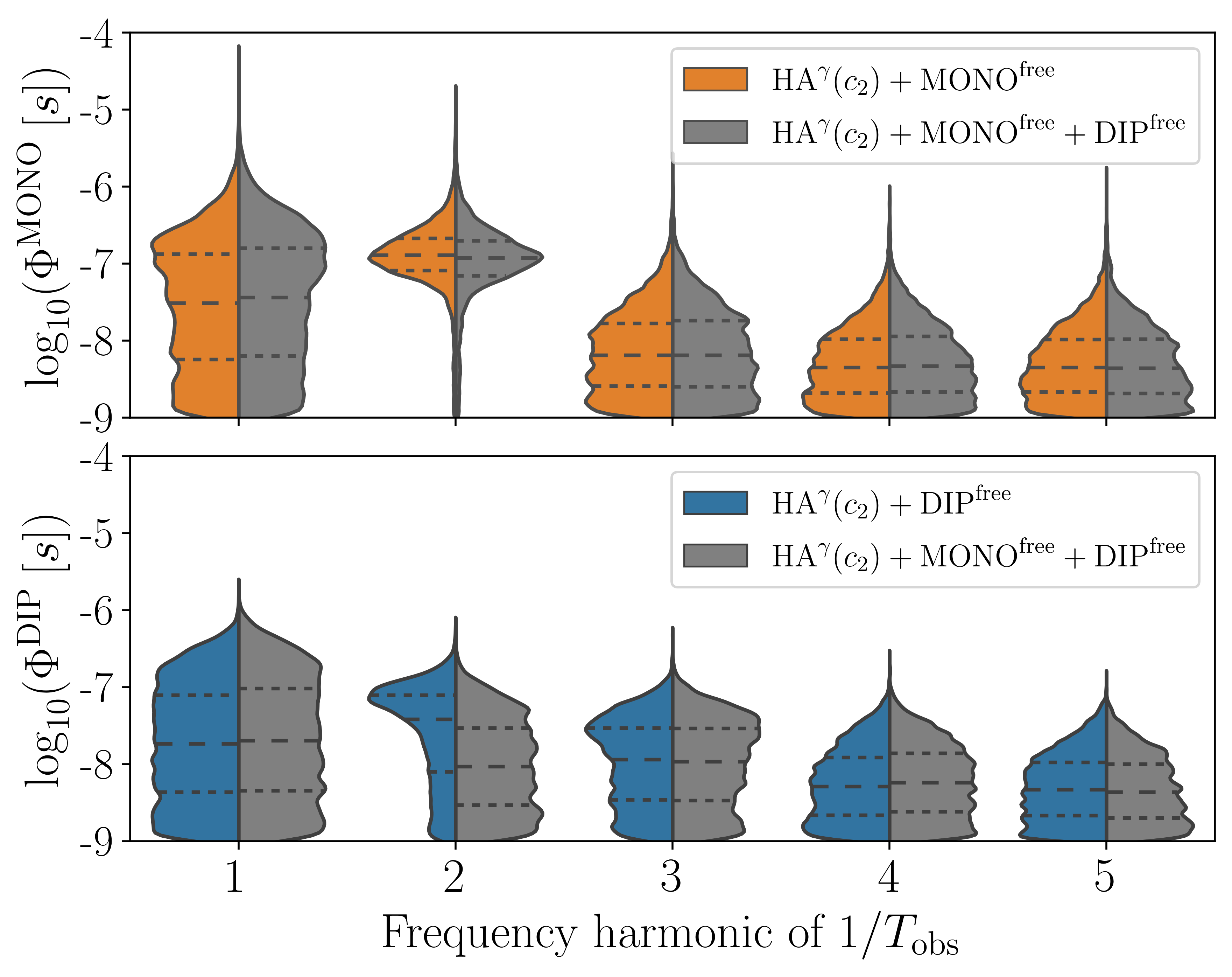}
    \caption{Marginalized 1D posterior distributions for the monopole (top plot) and dipole (bottom plot) free-spectrum parameters $\log_{10} \Phi_i$ of the first five frequency components $f_i = i/T_{\rm obs}$ where $i=1,...,5$.
    The green (right-hand) portion of the split violin plots show the results from a model that include both monopole and dipole free-spectrum models.
    For the \citetalias{15yrGWB} data set, $T_{\rm obs}=16.03$ years, which gives frequency components as multiples of $\approx \!\!2$~nHz.
    }
    \label{fig:monodip}
\end{figure}

\subsection{\label{subsec:analyses_monopole}Monopole and Dipole Correlations}

\begin{figure}[t]
    \centering
    \includegraphics[width =1\columnwidth]{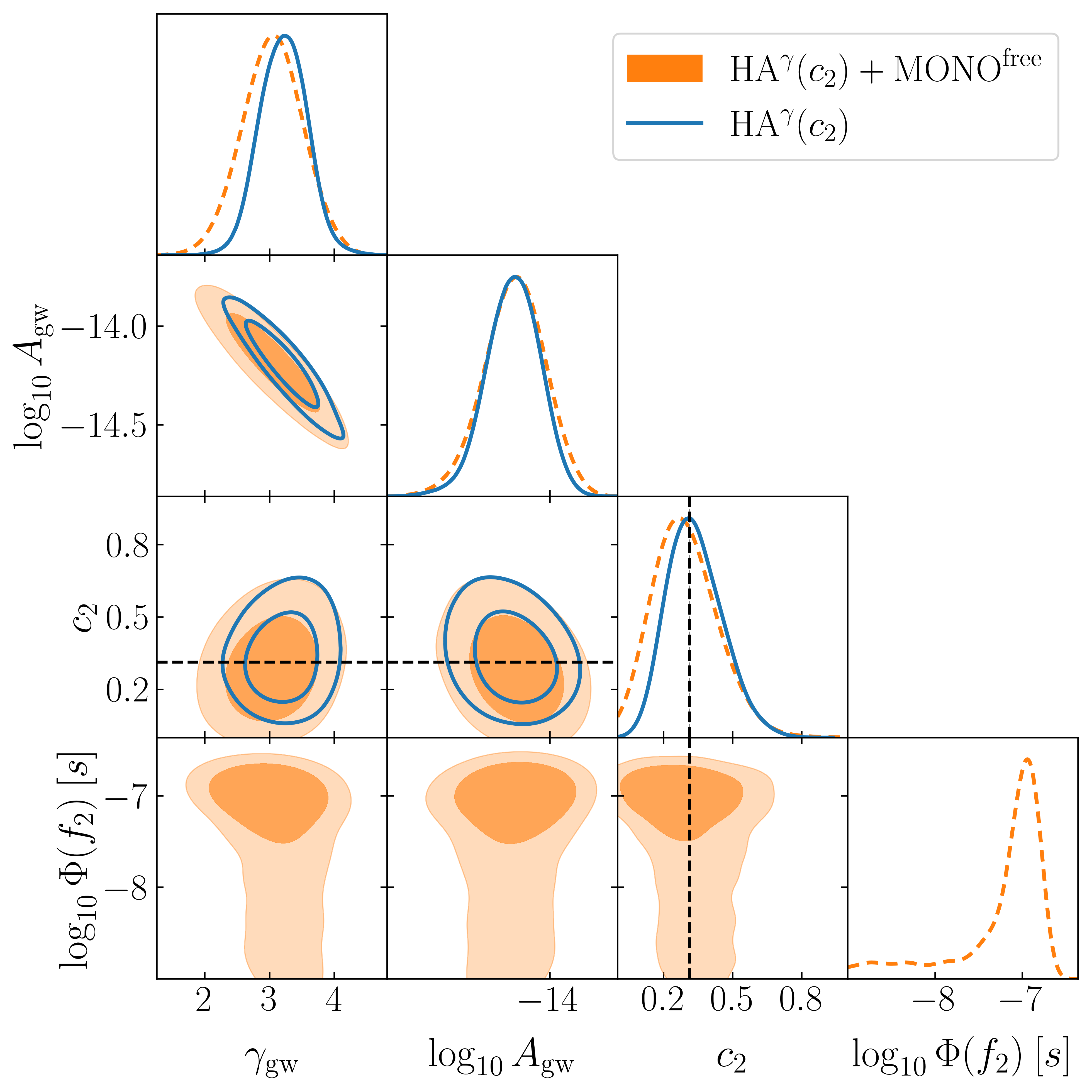}
    \caption{Marginalized 1D and 2D posterior distributions for a GWB analysis with monopole and quadrupole coefficients for the model $\text{HA}^\gamma(c_2)+\text{MONO}^{\rm free}$ (shown as orange regions and dotted-orange curves), overlaid with a quadrupole-only GWB analysis for the model $\text{HA}(c_2)$ (shown as solid blue curves).
    The monopole free-spectrum parameter $\log_{10} \Phi$ is associated with the second frequency bin ($f_2 = 2/T_{\rm obs} \approx 4 \; \text{nHz}$).
    The full model includes monopole free-spectrum parameters from the $1$st through $5$th frequency bins.
    The dashed black line shows the HD value of the quadrupole coefficient.
    }
    \label{fig:monocorr}
\end{figure}

To search for other possible correlations in the data, we add a monopole free-spectrum model ($\text{MONO}^{\rm free}$), dipole free-spectrum model ($\text{DIP}^{\rm free}$), or combination of both to GWB harmonic analysis model $\text{HA}^\gamma(c_2)$.
The violin plots in \autoref{fig:monodip} show the free-spectrum monopole and dipole models have power in the second frequency bin, $2/T_{\rm obs} \! \approx \! 4 \; \text{nHz}$.
We isolate this effect to the $\text{MONO}^{\rm free}$ model: even though a $\text{DIP}^{\rm free}$ model has some power in the second frequency bin, the $\text{MONO}^{\rm free}+\text{DIP}^{\rm free}$ model shows the data prefers monopolar power in the second harmonic and no dipolar power in any harmonic.
This result is consistent with other NANOGrav 15-year data set analysis results (e.g., see Figure~6 of \citetalias{15yrGWB}). 

\autoref{fig:monocorr} shows the marginalized 1D and 2D posterior distributions for model $\text{HA}^\gamma(c_2) + \text{MONO}^{\rm free}$. Note that although they are present in the analysis, we leave out all other frequency bins except the second, since this is the only bin that has a non-zero posterior distribution.  We overlay model $\text{HA}^\gamma(c_2)$ in \autoref{fig:monocorr} to show the effects on the harmonic analysis from including monopolar power in the model. This figure shows that when we include a monopole correlation, there is a reduction in the evidence for quadrupole correlations.
Specifically, the Savage-Dickey Bayes factor for parameter $c_2$ is reduced from 90 [model $\text{HA}^\gamma(c_2)$] to 5 [model $\text{HA}\gamma(c_2) + \text{MONO}^{\rm free}$].

\section{\label{sec:previous_work} Comparison with previous work}

Given the importance of the shape of the angular correlations, several methods have been used in the literature to extract angular information from the PTA data. The most common is to use a minimum variance estimator, known as the optimal statistic (OS), in order to either compute the angular correlations within a set of angular bins \citep{Anholm:2008wy,2013ApJ...762...94D,2015PhRvD..91d4048C,Vigeland:2018ipb}.  More recently, a generalization of the OS has been developed \citep{Sardesai:2023qsw}: the MCOS allows for multiple correlations to be simultaneously fit to the data. In \citetalias{15yrGWB} the MCOS is used to estimate the amplitude of a Legendre polynomial expansion of the angular correlation function. Finally, in both \citetalias{12.5yrGWB} and \citetalias{15yrGWB}, a splined angular correlation function is fit to the data, with knots placed over seven spline-knot positions \citep{Taylor:2012wv}. The choice of seven spline-knot positions is based on features of the HD curve.

The MCOS constraints on the amplitude of the Legendre polynomials is closest to the analyses presented here. As shown in Figure~7 of \citetalias{15yrGWB}, the amplitude for the quadrupole, $A_2 = A_{\rm gw}^2 c_2$, is significantly non-zero, and the monopole, $A_0 = A_{\rm gw}^2 c_0$, has a relatively small amplitude but is also significantly non-zero. However, the significance of these non-zero multipoles is notably larger than what we have found here. 

To understand this difference, we need to better understand the limitations of the MCOS analysis. The MCOS analysis presented in \citetalias{15yrGWB} provides a minimum variance estimator of a parameterization of the auto-correlations \emph{for a fixed model of the auto-correlations}, under the approximation of the \emph{weak signal limit} (where the inverse of the pulsar covariance is assumed to be dominated by intrinsic noise), and \emph{does not include correlations between different pairs of pulsars arising from the fact that they measure the same gravitational wave background}. 

The full uncertainty in the MCOS comes from two distinct sources: one is the variance in the estimator, and the other is the uncertainty in the parameter values that model the auto-correlations (such as $A_{\rm gw}$ and the intrinsic red noise parameters). A method to marginalize over the uncertainty in the auto-correlations was presented in \cite{Vigeland:2018ipb} and used to compute the posterior distributions in Figure~7 of \citetalias{15yrGWB}. However, the estimator uncertainty is of a similar size as the uncertainty from the marginalization. Recently, \cite{Gersbach:2024hcc} proposed ``uncertainty sampling'' as a way to consistently combine both of these sources of uncertainty, but we stress that the error bars in Figure~7 of \citetalias{15yrGWB} do not include the uncertainty in the estimator itself. 

In addition to this, the assumption of a weak signal already breaks down for the \citetalias{15yrGWB}. Thus, the uncertainty in the MCOS is inherently an underestimate of the true uncertainty in the shape of the angular cross-correlations \citep{Gersbach:2024hcc}. Neglecting the pulsar-pair cross-correlations can underestimate the uncertainty by $20-40\%$ \citep{15yrGWB}.

\begin{figure}[t]
    \centering
    \includegraphics[width =1\columnwidth]{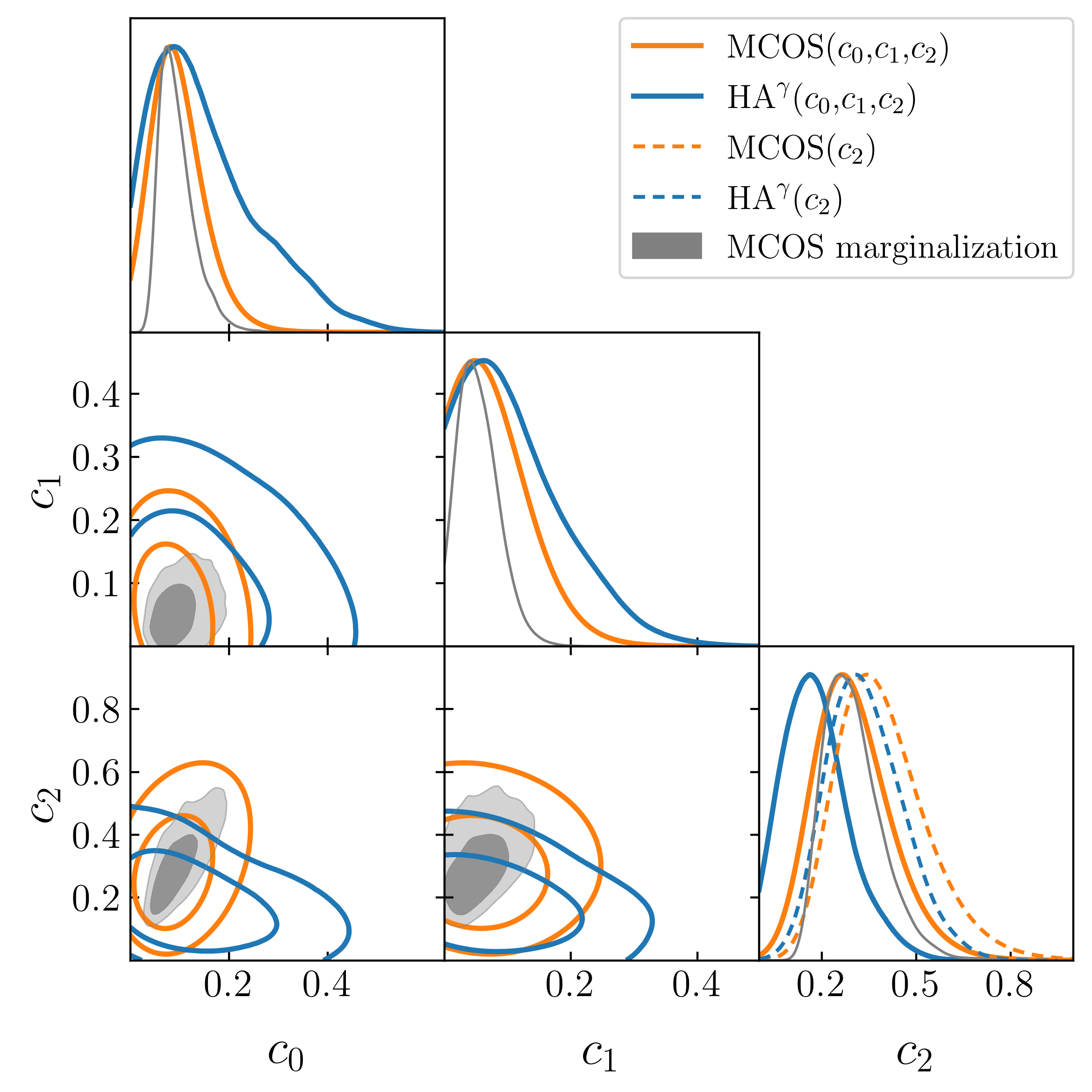}
    \caption{Marginalized 1D and 2D posterior distributions of frequentist MCOS and Bayesian harmonic analysis models with parameters $c_0$, $c_1$, and $c_2$.
    The gray curves show the MCOS analysis that appears in \citetalias{15yrGWB}. The uncertainty in this analysis solely comes from marginalizing over the ${\rm CURN}^\gamma$ parameters. The organge curves show the MCOS results when marginalizing over both the ${\rm CURN}^\gamma$ parameters and the estimator uncertainty, as outlined in \citep{Gersbach:2024hcc}.
    We do not include pulsar-pair covariance uncertainty for the MCOS results.
    The blue curves show the results of a harmonic analysis. 
    The additional dashed curves in the 1D marginalized constraints on $c_2$ show $c_2$-only analyses, as indicated in the figure legend. 
    }
    \label{fig:MCOS}
\end{figure}

To directly compare the MCOS results with our work, we show the results of an analysis that includes the monopole, $c_0$, dipole, $c_1$, and quadrupole, $c_2$, in the solid curves in \autoref{fig:MCOS}. It is important to note that the harmonic analysis approach requires Legendre coefficients to be positive (given that the angular power spectrum is positive) and the sum of Legendre coefficients to be less than 1 (to maintain a positive definite GWB covariance matrix), whereas there is no constraint on the prior range of the $c_\ell$ coefficients in the MCOS approach. We only show positive values for the MCOS results for simplicity. The gray lines and contours show the MCOS posteriors when only accounting for the uncertainty in the auto-correlations, whereas the orange lines and contours include both uncertainty in the auto-correlations as well as the estimator uncertainty, following the proscription outlined in \cite{Gersbach:2024hcc}. The blue contours show the result of the harmonic analysis presented here. In order to cast the MCOS results in terms of the multipoles, we re-scale the MCOS values by the value of $A^2_{\rm gw}$ obtained from the same point in the ${\rm CURN}^\gamma$ analysis chain. 

In \autoref{fig:MCOS}, we also overlay the MCOS and harmonic analysis marginalized 1D posterior distribution for a $c_2$-only model, shown as dashed curves. We see that the monopole coefficient has a larger impact on the posterior of the quadrupole for the harmonic analysis, compared to the MCOS approach. This is due to the fact that, as evident in the figure, the MCOS shows a positive correlation between $c_0$ and $c_2$, whereas the harmonic analysis shows a slight negative correlation. Intuitively, we expect the correlation to be negative: to keep the overall amplitude of the cross-correlations approximately constant, an increase in one of these coefficients would have to be accompanied by a decrease in the other. 

We stress that the harmonic analysis presented here does not suffer from any of the issues identified for the MCOS: the harmonic analysis directly utilizes the PTA likelihood and does not make any assumptions about the relative amplitude of the cross-correlations, and the posterior distributions automatically take into account all sources of uncertainty. 

\begin{figure}[t]
    \centering
    \includegraphics[width =1\columnwidth]{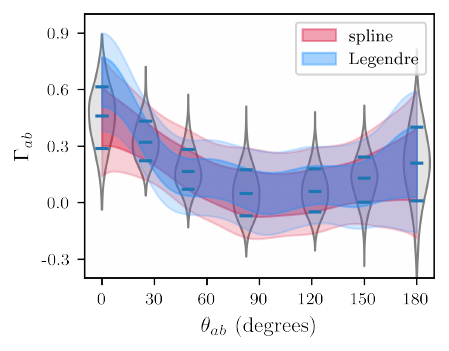}
    \caption{A comparison between constraints on the angular correlation function using a binned function \citep{Taylor:2012wv} (gray violins and red band) and the Legendre polynomial expansion ${\rm HA}^\gamma(c_0,\dots c_6)$ \citep{HAforPTAs}. The bands were constructed by drawing a large number of samples from the respective MCMCs and then computing the 1 and 2 $\sigma$ regions as a function of angular separation. The dashed black curve is the HD function. 
    }
    \label{fig:splineorf}
\end{figure}

Finally, we note that the MCMC spline analysis discussed in \S{}3 of \citetalias{15yrGWB} (and shown in \autoref{fig:splineorf}) provides similar information to our constraints on the multipoles. Just as in the harmonic analysis, the spline analysis fixes the correlation at $\theta_{ab} = 0$ to unity, effectively separating the auto-correlations from the cross-correlations \citep{Taylor:2012wv}. Since there are seven spline knots, we compare the spline analysis to the harmonic analysis ${\rm HA}^\gamma(c_0,\dots,c_6)$. In addition to having the same number of additional parameters, this harmonic analysis models variations in the correlations on the same angular scales. We show a comparison between these analyses in \autoref{fig:splineorf}. Both methods are consistent with one another and indicate that most of the evidence for non-zero angular-correlations comes from pulsar pairs separated by $\lesssim 30^\circ$. 

\section{\label{sec:discussion}Discussion and Conclusions} 

In this paper we extend the work of \citetalias{15yrGWB} to further characterize the angular correlations in the NANOGrav 15-year data set using the harmonic analysis method from~\cite{HAforPTAs}. Our results show the Bayesian evidence for quadrupole-only correlation models are consistent with the evidence for the HD model found in \citetalias{15yrGWB}. We do not see evidence for multipoles higher than the quadrupole. We find the HD value of the quadrupole coefficient falls within the 68\% CL of the measured quadrupole's marginalized 1D posterior distribution, $c_2/c_2^{\rm HD} = 1.088^{+0.32}_{-0.45}$. We reconstruct the angular correlation function from the posterior distributions for $c_2$ and $c_3$ and find the measurements are consistent with the angular correlations of the HD curve. 

We show the large jump in quadrupole evidence between the \citetalias{12.5yrGWB} and \citetalias{15yrGWB} data sets is primarily due to the increase in the observation time, which is consistent with a GWB power law that has increasing strength at lower frequencies. This result is consistent with the expected scaling of the mean-to-standard deviation ratio of the quadrupole with time versus number of pulsars found in \cite{HAforPTAs}. The current mean-to-standard deviation ratio of the octopole is $\sim 1$, and using the scaling found in \cite{HAforPTAs}, we expect to have a mean-to-standard deviation ratio $\simeq 3$ within roughly 10 years. 

Previous work in determining the shape of the measured angular correlations have either used a binned estimator of the correlations between pairs of pulsars (MCOS) or placed constraints on a spline parameterization of the angular correlations. Both of these approaches give broadly similar results to what we have found but also present some challenges when trying to interpret them. In particular, the MCOS results presented in \citetalias{15yrGWB} neglect cross-correlations between different pairs of pulsars, leading to an underestimate of the resulting uncertainty, and constraints on the amplitude of the knots of the spline analysis are highly correlated with one another. We note that recent work on improving the frequentest estimator has included all pulsar cross-correlations, but that this has yet to be applied to providing constraints on the shape of the angular correlations \citep{Gersbach:2024hcc}.  

When we include a monopole free-spectrum model in our harmonic analysis, the evidence of the quadrupole correlation is reduced by more than an order of magnitude due to the presence of a monopolar signal at $\approx\!\!4$~nHz, which has been investigated in detail.
Clock errors can cause monopole correlations~\citep{Tiburzi:2015kqa, 12.5yrGWB}, but extensive investigations by NANOGrav have shown no evidence of this type of systematic error (see discussion in \S{}5.3 of \citetalias{15yrGWB}).
Other possible explanations include GWs from an individual SMBHB source~\citep{NANOGrav:2023pdq}, ultralight dark matter~\citep{NG15new_physics}, and additional or alternate GW polarization modes~\citep{NANOGrav:2023ygs}.
To date, none of these additional investigations have been able to explain the source of the monopolar signal.
As stated in \citetalias{15yrGWB}, if this monopolar signal is due to an astrophysical or cosmological source, then its persistence in future data sets will help determine the source of the signal. We also note that, to date, it is unclear whether the other PTA data sets are consistent with this monopolar signal, and we leave such an analysis to future work. 

Our choice of parameterizing the angular correlations though Legendre polynomials is not unique---any set of complete functions can be used. For example it is possible to develop a set of functions which are statistically uncorrelated with the HD function (often referred to as ``principal components'')~\citep{Madison:2024fup}. An analysis that uses these functions may be better suited to more clearly identify potential deviations from the expected angular correlations. However, given that the expected angular power spectrum is predominantly quadrupolar, a Legendre polynomial expansion is the natural choice when assessing the Bayesian evidence for angular correlations. 

As our PTA data sets improve, it will be imperative to have multiple techniques to characterize the frequency and angular information contained within them. A confirmation of the standard expectations will lend credence to the interpretation that the GWB observed by PTAs is generated by a cosmological collection of SMBHBs generating tensor GWs that propagate at the speed of light. Even within this paradigm, we expect deviations due to the fact that the SMBHBs form a finite population that cluster on cosmological scales. Deviations from the standard expectations may provide evidence for unexpected dynamics in the SMBHB population, modifications to GR, and/or the presence of exotic GW sources such as cosmic strings or early universe phase transitions [see, e.g., \cite{NG15new_physics}]. The confirmation that the currently measured angular correlations in the \citetalias{15yrGWB} data set is largely consistent with standard expectations---though with hints of a possible monopole---is just a first step into a very exciting future. 

\section*{\label{subsec:acknowledgments}Acknowledgments}

\textit{Author contributions: }
An alphabetical-order author list was used for this paper in recognition of the fact that a large, decade timescale project such as NANOGrav is necessarily the result of the work of many people. All authors contributed to the activities of the NANOGrav collaboration leading to the work presented here, and reviewed the manuscript, text, and figures prior to the paper's submission. 
Additional specific contributions to this paper are as follows.

J.E.N. wrote and developed new python codes to perform the analysis, created figures and tables, and wrote a majority of the text. T.L.S. made significant contributions to the text, ran some of the analyses, and created some of the figures. 
K.K.B., T.L.S., and C.M.F.M. conceived of the project, supervised the analysis, helped write and develop the manuscript, and provided advice on figures and interpretation.
A.S. provided insights into the analysis, helped interpret the results, and provided comments on the manuscript.

G.A., A.A., A.M.A., Z.A., P.T.B., P.R.B., H.T.C., K.C., M.E.D., P.B.D., T.D., E.C.F., W.F., E.F., G.E.F., N.G., P.A.G., J.G., D.C.G., J.S.H., R.J.J., M.L.J., D.L.K., M.K., M.T.L., D.R.L., J.L., R.S.L., A.M., M.A.M., N.M., B.W.M., C.N., D.J.N., T.T.P., B.B.P.P., N.S.P., H.A.R., S.M.R., P.S.R., A.S., C.S., B.J.S., I.H.S., K.S., A.S., J.K.S., and H.M.W. ran observations and developed timing models for the NG15 data set.

\textit{Acknowledgments: }The work contained herein has been carried out by the NANOGrav collaboration, which receives support from the National Science Foundation (NSF) Physics Frontier Center award numbers 1430284 and 2020265, the Gordon and Betty Moore Foundation, NSF AccelNet award number 2114721, an NSERC Discovery Grant, and CIFAR. The Arecibo Observatory is a facility of the NSF operated under cooperative agreement (AST-1744119) by the University of Central Florida (UCF) in alliance with Universidad Ana G. M$\acute{\text{e}}$ndez (UAGM) and Yang Enterprises (YEI), Inc. The Green Bank Observatory is a facility of the NSF operated under cooperative agreement by Associated Universities, Inc.
K.K.B. and J.E.N. acknowledge the Texas Advanced Computing Center (TACC) at The University of Texas at Austin for providing high performance computing resources that contributed to the analyses reported within this paper.
T.L.S. acknowledges the Strelka Computing Cluster run by Swarthmore College that contributed to the analyses reported within this paper.
L.B. acknowledges support from the National Science Foundation under award AST-1909933 and from the Research Corporation for Science Advancement under Cottrell Scholar Award No. 27553.
P.R.B. is supported by the Science and Technology Facilities Council, grant number ST/W000946/1.
S.B. gratefully acknowledges the support of a Sloan Fellowship, and the support of NSF under award \#1815664.
The work of R.B., R.C., D.D., N.La., X.S., J.P.S., and J.A.T. is partly supported by the George and Hannah Bolinger Memorial Fund in the College of Science at Oregon State University.
M.C., P.P., and S.R.T. acknowledge support from NSF AST-2007993.
M.C. and N.S.P. were supported by the Vanderbilt Initiative in Data Intensive Astrophysics (VIDA) Fellowship.
K.Ch., A.D.J., and M.V. acknowledge support from the Caltech and Jet Propulsion Laboratory President's and Director's Research and Development Fund.
K.Ch. and A.D.J. acknowledge support from the Sloan Foundation.
Support for this work was provided by the NSF through the Grote Reber Fellowship Program administered by Associated Universities, Inc./National Radio Astronomy Observatory.
Support for H.T.C. is provided by NASA through the NASA Hubble Fellowship Program grant \#HST-HF2-51453.001 awarded by the Space Telescope Science Institute, which is operated by the Association of Universities for Research in Astronomy, Inc., for NASA, under contract NAS5-26555.
K.Cr. is supported by a UBC Four Year Fellowship (6456).
M.E.D. acknowledges support from the Naval Research Laboratory by NASA under contract S-15633Y.
T.D. and M.T.L. are supported by an NSF Astronomy and Astrophysics Grant (AAG) award number 2009468.
E.C.F. is supported by NASA under award number 80GSFC21M0002.
G.E.F., S.C.S., and S.J.V. are supported by NSF award PHY-2011772.
K.A.G. and S.R.T. acknowledge support from an NSF CAREER award \#2146016.
The Flatiron Institute is supported by the Simons Foundation.
S.H. is supported by the National Science Foundation Graduate Research Fellowship under Grant No. DGE-1745301.
N.La. acknowledges the support from Larry W. Martin and Joyce B. O'Neill Endowed Fellowship in the College of Science at Oregon State University.
Part of this research was carried out at the Jet Propulsion Laboratory, California Institute of Technology, under a contract with the National Aeronautics and Space Administration (80NM0018D0004).
D.R.L. and M.A.Mc. are supported by NSF \#1458952.
M.A.Mc. is supported by NSF \#2009425.
C.M.F.M. was supported in part by the National Science Foundation under Grants No. NSF PHY-1748958 and AST-2106552.
A.Mi. is supported by the Deutsche Forschungsgemeinschaft under Germany's Excellence Strategy - EXC 2121 Quantum Universe - 390833306.
P.N. acknowledges support from the BHI, funded by grants from the John Templeton Foundation and the Gordon and Betty Moore Foundation.
The Dunlap Institute is funded by an endowment established by the David Dunlap family and the University of Toronto.
K.D.O. was supported in part by NSF Grant No. 2207267.
T.T.P. acknowledges support from the Extragalactic Astrophysics Research Group at E\"{o}tv\"{o}s Lor\'{a}nd University, funded by the E\"{o}tv\"{o}s Lor\'{a}nd Research Network (ELKH), which was used during the development of this research.
S.M.R. and I.H.S. are CIFAR Fellows.
Portions of this work performed at NRL were supported by ONR 6.1 basic research funding.
J.S. is supported by an NSF Astronomy and Astrophysics Postdoctoral Fellowship under award AST-2202388, and acknowledges previous support by the NSF under award 1847938.
C.U. acknowledges support from BGU (Kreitman fellowship), and the Council for Higher Education and Israel Academy of Sciences and Humanities (Excellence fellowship).
C.A.W. acknowledges support from CIERA, the Adler Planetarium, and the Brinson Foundation through a CIERA-Adler postdoctoral fellowship.
O.Y. is supported by the National Science Foundation Graduate Research Fellowship under Grant No. DGE-2139292.

\bibliography{main}{}
\bibliographystyle{aasjournal}

\end{document}